\documentclass{article}

\usepackage{amssymb,amsmath,mathtools,xcolor,graphicx,xspace,colortbl,ragged2e,rotating} %
\usepackage{amsmath}  
\graphicspath{{resubmission_2 september 2021_graphics/}{resubmission_2 september 2021_tcache/}{resubmission_2 september 2021_gcache/}}
\DeclareGraphicsExtensions{.pdf,.eps,.ps,.png,.jpg,.jpeg}

\begin{document}
\title{Unobservable causal loops as a way to explain both the quantum computational speedup and quantum
nonlocality}
\author{
G. Castagnoli\protect\footnote{ Formerly: ICT Division and Quantum Optics Laboratory of Elsag Bailey, now part
of the Leonardo Company. Email: giuseppe.castagnoli@gmail.com
}}

\maketitle
\begin{abstract}We consider the reversible processes between two one-to-one correlated measurement outcomes which underly both problem-solving and quantum nonlocality. In the former case the two outcomes are the setting and the solution of the problem, in the latter those of measuring a pair of maximally entangled observables whose subsystems are space separate. We argue that the quantum description of these processes mathematically describes the correlation but leaves the causal structure that physically ensures it free, also of violating the time-symmetry required of the description of a reversible process. It would therefore be incomplete and could be completed by time-symmetrizing it. This is done by assuming that  the two measurements evenly contribute to selecting the pair of correlated measurement outcomes. Time-symmetrization leaves the ordinary quantum description unaltered but shows that it is the quantum superposition of unobservable time-symmetrization instances whose causal structure is completely defined. Each instance is a causal loop: causation goes from the initial to the final measurement outcome and then back from the final to the initial outcome. In the speedup, all is as if the problem solver knew in advance half of the information about the solution she will produce in the future and could use this knowledge to produce the solution with fewer computation steps. In nonlocality, the measurement on
either subsystem retrocausally and locally changes the state of both subsystems when the two were not yet spatially separate. This locally causes the correlation between the two future measurement outcomes.
\end{abstract}

\section{Introduction
}

The present work is a further step of the evolutionary approach $\left [1 -4\right ]$. In it, with others, we developed a retrocausal explanation of the quantum computational speedup still ad hoc for quantum computation. In the present work, we detach it from quantum computation and show that it derives from a more universal property of quantum correlation that also explains quantum nonlocality.

The potentially paradoxical character of the approach is clear from the abstract. The fact that the problem solver could benefit from the advanced knowledge of half of the information about the solution\protect\footnote{
For the time being let us think for simplicity that the solution is an invertible function of the problem-setting.
} she will produce and read in the future to produce the solution more efficiently
is reminiscent of the inventor
of the time machine sending back in time to herself, before she had the idea, the design of the machine. Although the latter is a favored theme of novels and movies, from a physical standpoint it is obviously nonsense.
The present explanation of the quantum speedup is, so to speak, half of
that nonsense. So, is it still nonsense or is it physically permissible? The present work deals with this question.

By the way, as quantum computation is still a specialized field and this work aims to be interdisciplinary, let us provide an example of quantum computational speedup, that of the simplest instance of Grover quantum search algorithm $[5]$. 

Bob, the problem-setter, hides a ball in a chest of four drawers. Alice, the problem-solver, is to identify the location of the ball by opening drawers. In the classical and worst case, to solve the problem Alice has to open three drawers: if the last drawer opened contains the ball, she has solved the problem; if not, it must be in the only drawer not yet opened and she has solved the problem as well. In the quantum case, i.e. by Grover algorithm, she always identifies the location of the ball by opening just one drawer. There is a quantum computational speedup. Also this might appear paradoxical of course. 

At the level of feeling, saying that, in the quantum case, Alice opens just one drawer but in a quantum superposition of the four drawer numbers might mitigate that impression. In the following we will try to put things together in an exact way.

Let us start from an only apparently remote lead. It is of course well known that Einstein, with Podolsky and Rosen $\left [6\right ]$, stated that the quantum description is incomplete because (in particular) it does not physically explain quantum nonlocality, namely the \textit{spooky action at a distance}. By the way, this particular case was the real concern of Einstein and is the one of interest here. 

In this work, we revamp that incompleteness issue starting from another lead. We consider the quantum description of quantum correlation (forgive the pun). We mean the reversible process between two one-to-one correlated measurement outcomes. Let us show in the first place that this process is common to the quantum speedup and quantum nonlocality.

In the speedup, the initial and final measurement outcomes are respectively the outcome of the initial measurement, which selects the problem-setting out of a uniform quantum superposition of all the possible problem-settings, and the corresponding solution of the problem selected by the final measurement (in the four drawer example, they are both the number of the drawer with the ball). The reversible process between the two outcomes is of course the quantum algorithm.

In nonlocality, the two outcomes are those of measuring two maximally entangled observables whose subsystems are space separate.

By the way, note that in both cases the process between the two measurement outcomes is reversible in spite of comprising the final measurement (which does not change the quantum state). In fact there is a unitary transformation between the two one-to-one correlated outcomes.

Now, our thesis is that the\textit{ ordinary quantum description}\protect\footnote{
We call the usual quantum description \textit{ordinary }to distinguish it form the complete quantum description that we are going to propose.
}  of quantum correlation is incomplete because it mathematically describes the correlation between the two measurement outcomes but not the causal structure that physically ensures it. Because of the reversibility of the quantum process, the process causal structure would be left free by the ordinary quantum description, also free of violating the time-symmetry required of the description of a reversible process. This would say, first, that this description is incomplete and, second, that it is completed by time-symmetrizing it. A basic assumption is that this requires evenly and non-redundantly sharing between the initial and the final measurement the selection of the information that specifies the sorted out pair of correlated measurement outcomes among all the possible pairs. This is of course an unorthodox assumption, we will justify it in a detailed way in the next section.
We should proceed as follows.

The initial measurement selects one of the possible halves of  the information in question. The respective measurement outcome unitarily propagates toward the final measurement until becoming the state immediately before it. The latter measurement selects the remaining half of the information. The respective measurement outcome unitarily propagates toward the initial measurement, by the inverse of the previous unitary transformation,  until becoming its definitive outcome. The latter propagation, which inherits both selections, is an instance of the time-symmetrized quantum process. Note that the causal structure of each time-symmetrization instance is completely defined and is a causal loop. Causality goes from the initial to the final measurement outcome and then back from the final to the initial measurement outcome. Since there is a plurality of possible instances, each corresponding to a way of taking half of the information in question, we should take their quantum superposition.

We will see that this time-symmetrization leaves the ordinary quantum description unaltered, as it should be since the description of a reversible process should be time-symmetric to begin with. But, at the same time, it shows that this description is a quantum superposition of unobservable time-symmetrization instances whose causal structure is completely defined. These instances are unobservables because of course they vanish in their quantum superposition, namely in the ordinary quantum description (which leaves the causal structure that ensures the correlation free).

For the fact of describing the causal structure that physically ensures the correlation between the two measurement outcomes, with respect to the ordinary quantum description which leaves this causal structure free, they describe the quantum process in a more complete way. In a way that immediately explains the quantum computational speedup and quantum nonlocality.

We will see that, in the speedup case, the outcome of the initial measurement encodes the problem to be solved by Alice and Alice's knowledge of the solution. The projection of the quantum state associated with the final measurement that selects the remaining half of the information, retrocausally, by the inverse of the time forward unitary transformation, changes it. This change reduces the computational complexity of the problem to be solved and correspondingly increases Alice's knowledge of the solution. The latter goes from complete ignorance of the solution to knowledge of half of the information that specifies it (among all the possible solutions). Therefore, everything is as if Alice knew in advance half of the information about the solution she will produce and read in the future and could use this knowledge to produce the solution more efficiently than in the classical case. Of course, one can see the causal loop.

The explanation in question is quantitative in character, given an oracle problem, it allows to compute the number of oracle queries needed to solve it in an optimal quantum way\protect\footnote{
We will explain in the technical part of the work what oracle problems and oracle queries are. For the time being we can see them just as problems and computation steps.
}. It is the minimum number logically (classically) needed to find the solution given the advanced knowledge of half of the information that specifies it. 

By the way, let us note that this is a completely technical consequence of a fundamental interpretation of quantum mechanics. It is a \textit{synthetic} solution, i.e. axiomatically derived from fundamental principles, of the so called \textit{quantum query complexity problem}.  It is the problem of finding whether the number of oracle queries needed to solve an oracle problem (forgive the pun) in an optimal quantum way is quadratic or exponential (or the like) in problem size. It is central to quantum computer science and still unsolved in the general case $\left [7 ,8\right ]$. The fundamental principle presently at play is that the causal structure of a time-reversible process does not introduce a preferred time-direction of causality.

In the nonlocality case, the time-symmetrization of the quantum process between the two measurement outcomes revamps, with an improvement, the retrocausal explanation of quantum nonlocality given by Costa de Beauregard in 1953 $\left .\right .\left [9\right ]$.

Let us lay the situation down. Say that at time $t_{0}$ two photons are generated by parametric down-conversion in the state of maximally entangled polarizations and diverging momenta. At time $t_{1}$, after their spatial separation, we measure the polarization of one photon. Of course this instantly changes the polarization state also of the space-separate photon. There is the spooky action at a distance. At time $t_{2}$, we measure, in the same basis, the polarization of the other photon. There is one-to-one correlation between the two measurement outcomes.

 In the first place, Costa de Beauregard noted that the unitary transformation that connects the two measurement outcomes, in a mathematically equivalent way, can go from $t_{1}$ to $t_{2}$ either directly or via $t_{0}$, when the two photons are not yet space-separate. In the latter case, the unitary transformation goes first backwards in time from $t_{1}$ to $t_{0}$ and then forward in time from $t_{0}$ to $t_{1}$ and eventually $t_{2}$. In the former case there is action at a distance, which is certainly unphysical. In the latter, everything is local and therefore physical provided that we consider retrocausality physical too.

His explanation assumes that, in causal order, the initial measurement only projects the reduced density operator of the polarization of the respective photon leaving that of the space-separate photon unaltered. Then this projection, on its way toward the final measurement via $t_{0}$, at time $t_{0}$  locally projects the polarization state of both photons. The resulting state, still on its way toward the final measurement, at the time of the initial measurement emulates action at a distance. In other words, the action from the first measurement outcome to the polarization state of the space separate photon would be local, first through the unitary transformation from $t_{1}$ to $t_{0}$ then through the unitary transformation from $t_{0}$ to $t_{1}$. But the sum of the two times, one negative and the other positive, is zero; this is what emulates action at a distance.

According to the present approach, this explanation would have two drawbacks:

i) Since the initial measurement alone selects the sorted out pair of measurement outcomes, it violates the time-symmetry required of the reversible process between the two outcomes.

ii) Correspondingly, it involves the (retrocausal) change of a past state of the ordinary quantum description, namely the state of maximal polarization entanglement of the two not yet space-separate photons.

We go now to the explanation provided by the present work. We should complete the ordinary quantum description of the reversible process that connects the two measurement outcomes via $t_{0}$ by time-symmetrizing it. In each time-symmetrization instance, in causal order, the projection of the quantum state due to either measurement (which selects one of the two halves of the information about the outcomes pair) only projects the reduced density operator of the polarization of the photon on which the measurement is performed, leaving that of the other photon unaltered. Then this projection, going toward the other measurement via $t_{0}$, at time $t_{0}$ locally projects the polarization state of both photons. As we will see, together, these two projections (one for each measurement) cause the one-to-one correlation between the two measurement outcomes. 

The differences with respect to Costa de Beauregard explanation are: 

i) There are no violations of the time-symmetry required of the reversible process between the two measurement outcomes.

ii) Correspondingly, no past quantum state of the ordinary quantum description is changed.

Costa de Beauregard explanation would be unphysical, the present explanation would be physical, although the basic idea is the same.

Another difference of course is that the present explanation is provided by the completion of the quantum description of quantum correlation, which also explains the quantum speedup. Costa de Beauregard's explanation is ad-hoc for quantum nonlocality.

Note that the causal loop implicit in the time-symmetrization procedure this time generates two causal loops, one for each measurement. Each loop is the measurement  that retrocausally changes the state of the two not yet space-separate subsystems in a way that in turn contributes to causing the correlation between the two measurement outcomes.

By the way, all the above would seem to answer Einstein and the other's call in $\left [6\right ]$. It shows: i) why the ordinary quantum description is incomplete -- it would be so because it allows causal structures that violate the time-symmetry required of a reversible process and ii) how to complete it -- by time-symmetrizing it. Our hidden (unobservable) variables, which complete the ordinary quantum description by telling the two photons, when they are not yet space separate, how to behave in the future measurements on them (as Einstein and the others required), are the time-directions of the causations between the two measurement outcomes in the unobservable time-symmetrization instances. 

While all the above would seem to exactly answer Einstein's expectations, it does so in a way that likely was not imagined at the time. It is of course an unorthodox way also at the present time. However it benefits, so to speak, of two paradigm shifts well successive to Einstein and the others' time: time-symmetric quantum mechanics $\left [10 -16\right ]$ and quantum computation $\left [17\right ]$, with the respective speedup $\left [18 ,19\right ]$. By the way, the present unorthodox approach, in particular, has been inspired by $\left [12\right ]$, about the non-sequential behavior of the wave function -- what we call here the ordinary quantum description.

We should eventually note that the quantum computational speedup and quantum nonlocality are the two known tasks that cannot be performed classically. By definition, the speedup is with respect to the classical algorithms, in the case of quantum nonlocality we are dealing with quantum cryptography based on Bell's theorem $\left [20\right ]$ whose security has no classical parallel. This work shows that they both rely on quantum causal loops. This suggests the conjecture that the present unobservable form of quantum retrocausality has to do with the very difference between quantum and classical mechanics.

Given the unconventional character of the work, before getting lost in the formalizations, we provide in the next section a very detailed line of reasoning. It goes from the diagnosis of the incompleteness of the ordinary description of quantum correlation to the prescription of the way of completing it. In particular, it comprises our best justification of all the unorthodox assumptions taken in the present work.

\section{Line of reasoning
}
We provide our line of reasoning, segmented as a numbered sequence of consequential steps. Since it is an unorthodox line of reasoning that goes against conventions, we have done our best to be precise.

 \medskip

i) Our starting argument is that the ordinary quantum description of the reversible processes between two one-to-one correlated measurement outcomes of course mathematically describes the correlation but leaves the causal structure that physically ensures it free. 

Naturally this goes against the convention that causality only goes forward in time. Our justification for doing without this convention in the case of a reversible quantum process is twofold. On the one side, the convention in question is not implied by the dynamic equations of the reversible quantum process but is an add-on to them. On the other, it is a convention that clashes with the very definition of reversible process. Let us recall it: a deterministic process is time-reversible if the time-reversed process satisfies the same dynamic equations as the original process. Of course, in the time-reversed process, causality goes backwards in time. Hence the clash.

 \medskip

ii) Anyway, the first critical step of the present line of reasoning is doing without that convention. Without it, the ordinary quantum description (the dynamic equations) of course does not provide any information about the causal structure that ensures the correlation between the two measurement outcome. It leaves this causal structure free.

In the first place, the initial measurement outcome can be considered the cause of the final measurement outcome or, time-symmetrically, the final measurement outcome the cause of the initial one. We should keep in mind that between the two one-to-one correlated measurement outcome there is a unitary transformation.

We note that, until now, choosing either time-direction of causality does not add any information to the ordinary quantum description. It is like reading it from left to right or right to left. The information about the quantum process is the same in either case.

What is important is that, once accepted that causality can also go backwards in time, in the quantum world the ambiguity about the time-direction of causality can go down to a deeper level. Causality can go forward or backwards in time between any two corresponding (correlated) parts of the two measurement outcomes and, independently, forward or backwards in time between the two complementary parts.

Let us exemplify this in the four drawer instance of Grover algorithm. Say that the number of the drawer with the ball selected by the initial measurement (out of the quantum superposition of the four drawer numbers), in binary notation is $01$. The corresponding solution, selected by the final measurement, is naturally $01$ too. The sorted out pair of correlated measurement outcomes is thus $\left (01 ,01\right )$. 

Under the convention that causality only goes forward in time, the initial measurement selects the number $01$ and this causes the number $01$ ``selected'' -- just read -- by the final measurement.  As we will soon see, without that convention, the ordinary quantum description allows to assume a different picture. The left digit of the initial measurement outcome can be the cause of the left digit of the final measurement outcome and the right digit of the final measurement outcome can be the cause of the right digit of the initial measurement outcome. Correspondingly, we should share the selection of the sorted out pair of correlated measurement outcomes between the initial and final measurements. The initial measurement selects the left digit of the pair, the final measurement the right one.

Note that the ordinary quantum description leaves us free to choose which amount of information is selected by the initial measurement and which by the final measurement. The information that specifies this choice, among all the possible choices, is information about the quantum process that the ordinary quantum description, which leaves this choice free, does not provide. Once accepted the fact that causality in the quantum world can also go backwards in time, this fact alone shows that the ordinary quantum description is incomplete. There is information about the quantum process that it does not provide.

\medskip

iii) The second critical step of the present line of reasoning, allowed by the previous one, is asking that the causal structure that ensures the correlation between the two measurement outcomes does not introduce a preferred time-direction of causality. In other words, we ask that the initial and final measurements evenly (and of course non-redundantly) contribute to select the pair of correlated measurement outcomes among all the possible pairs. In equivalent terms we require that each measurement selects one of the possible halves of the information that specifies the pair of outcomes among all the possible pairs and the other measurement selects the other half. Since there is a plurality of ways of taking half of the information, we should take their quantum superposition. At point vii) we will see how to put all things together.

This would seem to be the natural thing to do since we are dealing with a time-reversible process. However we can justify it in a stronger way, by comparing the case that the two measurements are one after the other with the case that they are simultaneous.

First, we note that it is indifferent to perform the two measurements simultaneously or not. In the nonlocality case, we can postpone the initial measurement to the time of the final measurement, or anticipate the latter to the time of the initial measurement. In the case of the speedup, the initial measurement that selects the problem-setting (e. g. the number of the drawer with the ball) can be postponed to the time of the final measurement (also here the respective observables commute). In this case, Alice performs the algorithm in a quantum superposition of all the possible problem-settings. Then, simultaneously, Bob's measurement selects the problem-setting and Alice's measurement the corresponding solution. In either the nonlocality or the speedup case, everything evidently remains the same: action at a distance and -- respectively -- the maximum speedup achievable by the quantum algorithm.

Now, in the case that the two measurements are simultaneous, the assumption that they evenly contribute to the selection of the sorted out pair of outcomes becomes mandatory. As we will see in the formalizations, in both the speedup and the nonlocality case we are dealing with two measurements that simultaneously select two maximally entangled observables. There is a perfect symmetry between them and no reason for which they should differently contribute to the selection.

In order that everything remains the same when the two measurements are not simultaneous, one has to keep the causal structure of the simultaneous case. In other words, when made subsequent to each other, the two measurements should continue to evenly contribute to the selection of the outcome pair. In equivalent terms, we should require that the causal structure that physically ensures the correlation between the two measurement outcomes does not introduce a preferred time-direction of causality.

iv) The adoption of this requirement implies that the causal structures allowed by the ordinary quantum description can violate it. Of course, with the causal structure free, the amount of information about the outcomes pair selected by one measurement can be different from that selected by the other measurement.  This has in turn the two following consequences.

\medskip

v) The ordinary quantum description of the processes in question is incomplete because it allows causal structures that violate the no preferred time-direction of causality requirement.

\medskip

vi) To complete the ordinary quantum description, it is sufficient to time-symmetrize it.

\medskip

vii) All the above already defines how to time-symmetrize the ordinary quantum description of the processes in question. We have already anticipated this in the introduction, We repeat it here in the example of the four drawers problem.

The initial measurement, performed in a quantum superposition the four possible numbers of the drawer with the ball (\textit{numbers} for short), selects one of the possible halves of  the information that specifies the number (among the four possible numbers). For example, it selects the value $0$ of the left digit of the number. The corresponding measurement outcome is therefore the superposition of the numbers $00$ and $01$. This superposition unitarily propagates toward the final measurement until becoming the state immediately before it (as a superposition of two tensor products, each a number beginning with $0$ and the corresponding solution -- that same number). The latter measurement selects the remaining half of the information (say the value $1$ of the right digit, so that the number selected by the two measurements is $01$). The respective measurement outcome unitarily propagates toward the initial measurement, by the inverse of the previous unitary transformation,  until becoming its definitive outcome. The latter propagation, which inherits both selections, is an instance of the time-symmetrized quantum process. 

Note that the causal structure of each time-symmetrization instance is completely defined and is a causal loop: causation goes from the initial to the final measurement outcome and then back from the final to the initial outcome. Since there is a plurality of possible instances, each corresponding to a way of taking half of the information in question, we should take their quantum superposition.

By the way, we should note that the causal structure of these instances, whose formation essentially relies on quantum superpositions, would not be possible in a classical version of the algorithm, even under the idealization that there are reversible classical processes. We could not divide the evolution of the two-digit number into two evolutions, one for the left and the other for the right digit, perfectly isolated from one another as required if the time-directions of causality are to be opposite in them. In fact, oracle queries must work in an inseparable way on the two digits together. Going deeper into this topic could be an interesting research prospect.

\medskip

viii) To complete our line of reasoning, we must anticipate another result that will come out from the formalizations. We will see that the superposition of all the time-symmetrization instances yields the ordinary quantum description of the process between the two measurement outcomes back again. 

Of course this is as it should be, because the quantum description of a reversible process should be time-symmetric to begin with, so it is right that it remains unchanged under its time-symmetrization.

Also note that such instances, with a completely defined causal structure, are unobservable because they vanish in the superposition of all of them, which is indeed the ordinary quantum description which leaves the causal structure that physically ensures the correlation between the two measurement outcomes free.

\medskip               \medskip

ix) However, the fact that the time-symmetrization of the ordinary quantum description gives us the same description back again does not mean that we are left with nothing in hand, on the contrary. We have found that the ordinary quantum description of the reversible process between two one-to-one correlated measurement outcomes, which leaves the causal structure that physically ensures the correlation free, is a quantum superposition of unobservable time-symmetrization instances whose causal structure is completely defined.

\medskip

x) The key feature of these instances is that they are causal loops.

\medskip
\medskip

xi) Because of their completely defined causal structure, such causal loops describe the quantum process between the two measurement outcomes in a more complete way than their quantum superposition -- the ordinary quantum description which leaves the causal structure free.

\medskip

xii) As we will see in the formalizations, they immediately and exactly explain the quantum computational speedup and quantum nonlocality.

\medskip

xiii) Of course what completes the ordinary quantum description is the description of the quantum superposition it is made of, in fact that of the unobservable quantum causal loops.

\smallskip

With this, we have finished exposing our line of reasoning. What emerges seems to be a very interesting form of quantum retrocausality. 

Probably, the reason one is skeptical about the existence of retrocausality is its ability to change the past. This, in the macroscopic world, leads to well known paradoxical consequences. However the unobservable form of quantum retrocausality we are dealing with never changes the past as described by the ordinary quantum description, given that it vanishes in it. It should therefore be immune to the paradoxes that plague classical retrocausality.

\section{ Time-symmetrizing the ordinary quantum description of the computational speedup
}
This section is a review of the explanation of the quantum computational
speedup developed in the evolutionary approach
$\left [1 -4\right ]$, with further clarifications and in a form that immediately fits the notion of quantum causal loop.

Before proceeding, let us explain what an \textit{oracle}, an \textit{oracle}\textit{ problem,} and an \textit{oracle query} are in the present context. An oracle is a black box that computes a function, for
example the Kronecker delta. The oracle problem is finding, by means of the oracle, a characteristic of that function, for
example the argument for which the Kronecker delta is
$1$. Alice, the problem solver, gives the oracle a value of the argument of the function; in
jargon, she performs an \textit{oracle query}. The oracle gives her back the corresponding
value of the function. The all procedure counts for a single computation step, or oracle query. The
number of oracle queries needed to solve an oracle problem in an optimal quantum way is its \textit{quantum query complexit}y.

 It is natural to think that, in order to understand what happens within a quantum
process, one must have a description of the same as complete as possible. With respect to the usual
description of quantum algorithms, which is limited to the process of solving the problem, we physically describe also: (1) the process of setting the problem, (2) the
fact that consequently the problem-setting should be hidden from the problem solver -- otherwise it would tell her
the solution of the problem before she begins her problem-solving action. By the way, this is particularly evident in the four drawer problem, where the problem-setting is the number of the drawer with the ball which is also the solution of the problem. Only after these two completions of the object of the quantum description, we will complete the quantum description itself by time-symmetrizing it  -- this is step (3).

\subsection{Step 1 -- extending the usual quantum description to the process of setting the
problem
}
 Let us consider the simplest
instance of the oracle problem solved by Grover quantum search algorithm
$\left [5\right ]$
-- Grover's problem from now on. Bob hides a ball in one of four drawers.
Alice is to locate it by opening drawers. Opening drawer number
$a$
amounts to querying the oracle with the question: is the ball in
drawer
$a$? The oracle computes the Kronecker function
$\delta \left (b ,a\right )$, where
$b$
is the number of the drawer with the ball, and of course gives back
$1$
if
$a =b$
and
$0$
otherwise. As well known, the four drawer instance of Grover algorithm always yields the
solution of the problem -- locates the ball -- with just one oracle query, performed in a quantum
superposition of the four possible drawer numbers. Since in the classical case it may be necessary to
perform up to three queries, there is a
\textit{quantum
computational speedup}.

The usual description
of quantum algorithms is limited to Alice's problem-solving
action. We extend it to Bob's action of setting the problem.

Let us introduce the notation first:

 We number the four drawers in binary
notation:
$00 ,01 ,10 ,11$.

A quantum register
$B$, under the control of the problem setter Bob, is meant to contain the \textit{problem-setting }
$b\text{}$
(the number of the drawer with the ball). Its basis vectors are thus:
$\vert 00 \rangle _{B} ,\vert 01 \rangle _{B} ,\vert 10 \rangle _{B} ,\vert 11 \rangle _{B}$.

A quantum register
$A$, under the control of the problem solver Alice, is meant to contain
$a$, the number of the drawer that Alice wants to open; it will eventually contain the solution of the problem -- the number of the drawer with
the ball
$b$. Its basis vectors are thus:
$\vert 00 \rangle _{A} ,\vert 01 \rangle _{A} ,\vert 10 \rangle _{A} ,\vert 11 \rangle _{A}$.

 Let the observable
$\hat{B}$, of eigenstates/eigenvalues respectively
$\vert 00 \rangle _{B} ,\vert 01 \rangle _{B} ,\vert 10 \rangle _{B} ,\vert 11 \rangle _{B}$
/
$00 ,01 ,10 ,11$, be the number contained in register
$B$; the observable
$\hat{A}$, of eigenstates/eigenvalues respectively
$\vert 00 \rangle _{A} ,\vert 01 \rangle _{A} ,\vert 10 \rangle _{A} ,\vert 11 \rangle _{A}$
/
$00 ,01 ,10 ,11$, be the number contained in register
$A$.

 The process of setting and solving the
problem is represented by the following table (from now on we disregard normalization):

\begin{equation}\begin{array}{ccc}\begin{array}{c}\;\text{time }t_{1}\text{, meas. of}\;\hat{B}\end{array} & t_{1} \rightarrow t_{2} & \text{time }t_{2}\text{, meas. of}\;\hat{A} \\
\, & \, & \, \\
\left (\vert 00 \rangle _{B} +\vert  01 \rangle _{B} +\vert 10 \rangle _{B} +\vert  11 \rangle _{B}\right )\vert 00 \rangle _{A} & \, & \, \\
\Downarrow  & \, & \, \\
\vert 01 \rangle _{B}\vert  00 \rangle _{A} &  \Rightarrow \hat{\mathbf{U}}_{1 ,2} \Rightarrow  & \vert 01 \rangle _{B}\vert  01 \rangle _{A}\end{array} \label{ext}
\end{equation}

We assume that, in the initial state, the
number of the drawer with the ball is completely indeterminate. This is represented by a
uniform quantum superposition of all the basis vectors of register
$B$
-- top-left corner of the table (by the way, the basis of this register will never be rotated). Register
$A$
is initially in an arbitrary sharp state standing for a blank blackboard.

At time
$t_{1}$, Bob measures
$\hat{B}$
in that initial state. This projects the initial superposition on an eigenstate of
$\hat{B}$
selected at random, say
$\vert 01 \rangle _{B}$
(follow the vertical arrow on the left of the table). The corresponding eigenvalue,
$b =01$, is the problem-setting selected by Bob -- here the number of the drawer with the ball.
By the way, Bob could unitarily change it into a desired number. For simplicity we disregard this
operation.

We denote by
$\hat{\mathbf{U}}_{1 ,2}$
the unitary part of Alice's problem-solving action. It should send the input state
$\vert 01 \rangle _{B}\vert  00 \rangle _{A}$
into the output state
$\vert 01 \rangle _{B}\vert 01 \rangle _{A} =\hat{\mathbf{U}}_{1 ,2}\vert 01 \rangle _{B}\vert 00 \rangle _{A}$, which contains in register
$A$ the solution of the problem -- again the number of the drawer with the ball selected by Bob
-- right looking horizontal arrows.

Of course,
$\hat{\mathbf{U}}_{1 ,2}$
can be the unitary part of Grover algorithm, or even a less efficient quantum algorithm
that solves Grover's problem. However, to us, it can remain an unknown unitary transformation. We
only need to know that there can be a unitary transformation between the input
and output states, which is always the case with this kind of representation where the output has a
full memory of the input. Note that the input and output states of
$\hat{\mathbf{U}}_{1 ,2}$
are known once the oracle problem -- the set of the problem-settings and the
corresponding solutions -- is known. In other words, that of table (\ref{ext}) is an external description of
the quantum algorithm -- of what the quantum algorithm should do to solve the oracle problem -- which only depends on the oracle problem.
$\hat{\mathbf{U}}_{1 ,2}$, the unitary part of the quantum algorithm, can remain an unknown unitary transformation.

Eventually, at time
$t_{2}$, Alice acquires the solution by measuring
$\hat{A}$. The output state of
$\hat{\mathbf{U}}_{1 ,2}$, namely
$\vert 01 \rangle _{B}\vert 01 \rangle _{A}$, with register
$A$
already             in an eigenstate of
$\hat{A}$, remains unaltered; there is then a unitary transformation between the initial and
final measurement outcomes; the process between them is physically reversible as no information is
lost in it.

\subsection{ Step 2 -- relativizing the extended quantum description to the problem solver
}
Register
$B$
contains the problem-setting
$b$. Its state represents the knowledge
of the problem-setting on the part of Bob and any external observer. Initially it represents complete ignorance of it -- see the top-left corner
of table (\ref{ext}). After the initial measurement of $\hat{B}$ on the part of Bob, it represents full knowledge of it -- bottom-left corner of the table.

Therefore we can see that the description of
table (\ref{ext}) works for Bob and any external observer, not for
Alice. Immediately after the initial measurement, she would know the solution of the problem\protect\footnote{
This is particularly evident in the present case, where the problem-setting and the solution are both the number of the drawer with the ball. But it holds in general, since Alice always knows the function computed by the black box.
}
before she begins her problem-solving action. To Alice, the problem-setting
selected by Bob must be hidden inside a black box.

To physically represent the concealment of
the outcome of the initial measurement from Alice (the problem solver and observer of the final measurement), in the first place we must be able to have two different descriptions of the quantum algorithm, one with respect to Bob and any external observer and the other with respect to Alice. For this, we resort to relational quantum mechanics
$\left [21 ,22\right ]$. In it, quantum states are not absolute but relative to the observer. We\textit{
relativize} the description
of the quantum algorithm with respect to Alice as follows.

To her, we postpone after the end of her
problem-solving action the projection of the quantum state associated with the initial measurement
(note that
$\hat{B}$
and
$\hat{A}$
commute). This is of course a mathematically legitimate operation provided that the two
ends of the projection undergo the unitary transformation
$\hat{\mathbf{U}}_{1 ,2}$
\protect\footnote{ Alternatively, we could postpone at the end of Alice's action the very
measurement of
$\hat{B}$. This can be done because the reduced density operator of register
$B$
remains unaltered throughout
$\hat{\mathbf{U}}_{1 ,2}$. Note that, with reference to point iii) of Section 2, this would make the two measurements simultaneous. 
}.

While the description
of the quantum algorithm with respect to Bob and any external observer is that of table (\ref{ext}), that with respect to Alice becomes:

\begin{equation}\begin{array}{ccc}\begin{array}{c}\text{time}\text{ }t_{1}\text{, meas. of}\;\hat{B}\end{array} & t_{1} \rightarrow t_{2} & \text{\text{time }\text{}}t_{2}\text{, meas. of}\;\hat{A} \\
\, & \, & \, \\
\left (\vert 00 \rangle _{B} +\vert  01 \rangle _{B} +\vert 10 \rangle _{B} +\vert  11 \rangle _{B}\right )\vert 00 \rangle _{A} &  \Rightarrow \hat{\mathbf{U}}_{1 ,2} \Rightarrow  & \vert 00 \rangle _{B}\vert  00 \rangle _{A} +\vert 01 \rangle _{B}\vert  01 \rangle _{A} +\vert 10 \rangle _{B}\vert  10 \rangle _{A} +\vert 11 \rangle _{B}\vert  11 \rangle _{A} \\
\, & \, & \Downarrow  \\
\, & \, & \vert 01 \rangle _{B}\vert  01 \rangle _{A}\end{array} \label{alice}
\end{equation}

To Alice, the initial quantum superposition of the four possible numbers of the drawer with the ball
remains unaltered after the initial Bob's measurement of
$\hat{B}$
-- top-left corner of the table. This superposition represents
her complete ignorance of the number of the drawer with the ball selected by Bob. By
$\hat{\mathbf{U}}_{1 ,2}$
(horizontal arrows), this superposition unitarily evolves into the superposition of four
tensor products, each the product of a number of the drawer with the ball and the corresponding
solution (that same number but in register
$A$). Eventually, the final measurement of
$\hat{A}$
projects the latter superposition on
$\vert 01 \rangle _{B}\vert  01 \rangle _{A}$, the tensor product of the number of the drawer with the ball already selected by Bob
and the corresponding solution (vertical arrow). Note that this projection is indifferently that due
to the initial Bob's measurement, postponed.

\subsection{ Step 3 -- Completing the ordinary quantum description by time-symmetrizing it
}
With probability one of reading the solution,
there is the unitary transformation
$\hat{\mathbf{U}}_{1 ,2}$
between the initial and final measurement outcomes. The process between them
is reversible in spite of comprising the final measurement -- which does not change the quantum state.
Correspondingly, its ordinary quantum description leaves the direction of causality unspecified. We have seen that this allows causal structures that violate the time-symmetry required of the description of a reversible process (violate the requirement of the absence of a preferred time-direction of causality), thus showing that the ordinary quantum description of the process in question is incomplete. This also says that the description is completed by time-symmetrizing it.

 We
time-symmetrize the description by imposing that the initial and final measurements evenly
contribute to the selection of the information that specifies the sorted out pair of correlated
measurement outcomes (among all the possible pairs). Since there is a plurality of ways of sharing the information in question into two halves, we should take their quantum superposition.

We should assume that the initial measurement
of
$\hat{B}$
and the final measurement of
$\hat{A}$
in all the possible ways           reduce to complementary partial measurements that
evenly and non-redundantly contribute to the selection of the pair of outcomes.

 For example, the initial measurement of
$\hat{B}$
could reduce to that of
$\hat{B}_{l}$
(the left binary digit -- bit -- of the number in register
$B$) and the final measurement of
$\hat{A}$
to that of
$\hat{A_{r}}$
(the right bit of the number in register
$A$). Or vice-versa, etc. Together, the measurement of
$\hat{B}_{l}$
at time
$t_{1}$
and that of
$\hat{A_{r}}$
at time
$t_{2}$
select a well defined problem-setting, say it is
$01$. How to arrange together the selections performed by the two successive measurements in order that, together, they determine the process between the two outcomes is
explained in the next subsection.

\subsubsection{ Time-symmetrizing the description of the quantum algorithm
to Bob and any external observer
}
The description of the quantum algorithm
of table (\ref{ext})
is that with respect to Bob and any external observer. We time-symmetrize
it by the following\textit{ time-symmetrization (TS) formalism}:

\begin{equation}\begin{array}{ccc}\begin{array}{c}\;\text{time }t_{1}\text{, }\text{}\text{meas. of}\;\hat{B}_{l}\end{array} & t_{1 \rightleftarrows }t_{2} & \text{time }\text{}t_{2}\text{, meas. of}\;\hat{A_{r}} \\
\, & \, & \, \\
\left (\vert 00 \rangle _{B} +\vert  01 \rangle _{B} +\vert 10 \rangle _{B} +\vert  11 \rangle _{B}\right )\vert 00 \rangle _{A} & \, & \, \\
\Downarrow  & \, & \, \\
\left (\vert 00 \rangle _{B} +\vert  01 \rangle _{B}\right )\vert 00 \rangle _{A} &  \Rightarrow \hat{\mathbf{U}}_{1 ,2} \Rightarrow  & \vert 00 \rangle _{B}\vert  00 \rangle _{A} +\vert 01 \rangle _{B}\vert  01 \rangle _{A} \\
\, & \, & \Downarrow  \\
\vert 01 \rangle _{B}\vert  00 \rangle _{A} &  \Leftarrow \hat{\mathbf{U}}_{1 ,2}^{\dag } \Leftarrow  & \vert 01 \rangle _{B}\vert  01 \rangle _{A}\end{array} \label{sybo}
\end{equation}

The initial measurement of
$\hat{B}_{l}$, selecting the
$0$
of
$01$, projects the initial quantum superposition on the superposition of the terms
beginning with
$0$
-- vertical arrow on the left of the table. Under
$\hat{\mathbf{U}}_{1 ,2}$, the latter superposition evolves into the superposition of the two products
\textit{number of the drawer with the ball}
$ \otimes $
\textit{corresponding solution} -- right looking horizontal arrows. Then the
final measurement of
$\hat{A_{r}}$, selecting the
$1$
of
$01$, projects the superposition in question on the term ending in
$1$
(vertical arrow on the right of the table). The backwards in time propagation, by
$\hat{\mathbf{U}}_{1 ,2}^{\dag }$, of
this output state, which inherits both selections, is an instance of the time-symmetrized description
of the quantum algorithm with respect to Bob and any external observer -- bottom line of the table.
Moreover, it is also the superposition of all instances since their bottom lines are all identical
to each other. One can see that the bottom line of the table remains the same regardless of
how we evenly share the selection of the sorted out pair of measurement oucomes between the initial and final measurements. It is also identical
to the bottom line of table (\ref{ext}), namely to the ordinary
description
of the quantum algorithm, since the symbol ``$ \Leftarrow \hat{\mathbf{U}}_{1 ,2}^{\dag } \Leftarrow $'' is of course equivalent to ``$ \Rightarrow \hat{\mathbf{U}}_{1 ,2} \Rightarrow $''. As anticipated in the Introduction, the superposition of all the time-symmetrization
instances gives the ordinary quantum description back again -- in a trivial way in the present case.

The above zigzag diagram yields the time-symmetric causal structure that ensures the correlation between the two measurement outcomes in the case of the quantum algorithm with respect to the problem setter Bob and any external observer. We have given this for completeness. It will be the causal structure of the quantum algorithm with respect to Alice the one that immediately explains the quantum speedup.

By the way, with reference to point ii) of Section 2, one can see the essential role played by quantum superposition in order that the left digit $0$ of the number of the drawer with the ball $01$ selected by the initial measurement causes the corresponding digit of the final measurement outcome and the left digit $1$  selected by the final measurement causes the corresponding digit of the initial measurement outcome.

\subsubsection{
Time-symmetrizing the description of the quantum algorithm
to Alice
}
 We time-symmetrize the description
of the quantum algorithm with respect to Alice of table (\ref{alice}). The following zigzag diagram yields the time-symmetrization
instance for problem-setting
$b =01$, the measurement of
$\hat{B}$
reduced to that of
$\hat{B}_{l}$, and the measurement of
$\hat{A}$
to that of
$\hat{A_{r}}$:

\begin{equation}\begin{array}{ccc}\begin{array}{c}\;\text{time }t_{1}\text{, meas. of}\;\hat{B}_{l}\end{array} & t_{1 \rightleftarrows }t_{2} & \text{time }t_{2}\text{, meas. of}\;\hat{A_{r}} \\
\, & \, & \, \\
\left (\vert 00 \rangle _{B} +\vert  01 \rangle _{B} +\vert 10 \rangle _{B} +\vert  11 \rangle _{B}\right )\vert 00 \rangle _{A} &  \Rightarrow \hat{\mathbf{U}}_{1 ,2} \Rightarrow  & \vert 00 \rangle _{B}\vert  00 \rangle _{A} +\vert 01 \rangle _{B}\vert  01 \rangle _{A} +\vert 10 \rangle _{B}\vert  10 \rangle _{A} +\vert 11 \rangle _{B}\vert  11 \rangle _{A} \\
\, & \, & \Downarrow  \\
\left (\vert 01 \rangle _{B} +\vert  11 \rangle _{B}\right )\vert 00 \rangle _{A} &  \Leftarrow \hat{\mathbf{U}}_{1 ,2}^{\dag } \Leftarrow  & \vert 01 \rangle _{B}\vert  01 \rangle _{A} +\vert 11 \rangle _{B}\vert  11 \rangle _{A}\end{array} \label{syal}
\end{equation}
The projection of the quantum state associated with the initial measurement of
$\hat{B}_{l}$
must be postponed after the end of Alice's problem-solving action -- outside table
(\ref{syal}) which
is limited to this action. In fact any information about the problem-setting must be hidden from
her. The top line of the diagram is thus the same of table (\ref{alice}). Then the measurement of
$\hat{A_{r}}$
in the output state of
$\hat{\mathbf{U}}_{1 ,2}$, selecting the
$1$
of
$01$, projects it on the superposition of the terms ending with
$1$
(vertical arrow). The propagation of the latter superposition -- which inherits both selections -- backwards in time by
$\hat{\mathbf{U}}_{1 ,2}^{\dag }$
(left looking horizontal arrows) is an instance of the
time-symmetrized quantum algorithm to Alice. See the bottom line of table
(\ref{syal}). 

In view of what will follow, note that we can speak as well of projection of the top line of the diagram on the bottom line. Let us repeat this bottom line -- i.e. the time symmetrized quantum algorithm -- below for convenience:

\begin{equation}\begin{array}{ccc}\text{time }t_{1} & t_{1} \leftarrow t_{2} & \text{time }t_{2} \\
\, & \, & \, \\
(\vert 01 \rangle _{B} +\vert  11 \rangle _{B})\vert 00 \rangle _{A} &  \Leftarrow \hat{\mathbf{U}}_{1 ,2}^{\dag } \Leftarrow  & \vert 01 \rangle _{B} \vert 01 \rangle _{A} +\vert  11 \rangle _{B}\vert 11 \rangle _{A}\end{array} \label{distance}
\end{equation}

The above table can also be read the other way
around as:

\begin{equation}\begin{array}{ccc}\text{time }t_{1} & t_{1} \rightarrow t_{2} & \text{time }t_{2} \\
\, & \, & \, \\
(\vert 01 \rangle _{B} +\vert  11 \rangle _{B})\vert 00 \rangle _{A} &  \Rightarrow \hat{\mathbf{U}}_{1 ,2} \Rightarrow  & \vert 01 \rangle _{B} \vert 01 \rangle _{A} +\vert  11 \rangle _{B}\vert 11 \rangle _{A}\end{array} \label{disr}
\end{equation}

By the way, let us say for completeness
that the output state
$\vert 01 \rangle _{B}\vert  01 \rangle _{A} +\vert 11 \rangle _{B}\vert  11 \rangle _{A}$
will eventually be projected on
$\vert 01 \rangle _{B}\vert  01 \rangle _{A}$
by the projection due to the initial measurement of
$\hat{B}_{l}$
postponed after the end of Alice's action, outside table (\ref{syal}) which is limited to this action.

For
$b =01$, there are in total three time-symmetrization instances. In each of them, the
problem-setting selected by Bob,
$b =01$, pairs with another problem-setting (it pairs with
$b =11$
in the above instance). The superposition of all instances of the time-symmetrized quantum algorithm (one is that of table \ref{disr}), also for all the possible
problem-settings\protect\footnote{ In the quantum algorithm to Alice, the problem-setting is selected only
after the unitary part of Alice's problem-solving action.
}, gives back the unitary part of the original
quantum algorithm to Alice -- the top line of table (\ref{alice}). Of course, the
superposition of all the pairs of basis vectors of register
$B$
yields the superposition of all the basis vectors, namely the top-left state of table (\ref{alice}).

\subsubsection{ Interpretation
}
Table (\ref{distance}), or identically (\ref{disr}), yields an instance
of the time-symmetrized quantum algorithm with respect to Alice.
From now on we will refer to the latter table which is a more familiar description of a quantum
algorithm.

Note that, besides the element of a quantum
superposition, each instance can be seen itself as a full fledged quantum algorithm. In the instance
we are dealing with, it suffices to think that actually Bob measures
$\hat{B}_{l}$
in the initial state and Alice
$\hat{A_{r}}$
in the output state of
$\hat{\mathbf{U}}_{1 ,2}$
-- of course this selects the entire problem-setting $01$ too. The time-symmetrization of the quantum algorithm in this case only
consists in propagating forward in time the selection due to the initial measurement and backwards
in time that due to the final measurement. The time-symmetrized quantum algorithm in this case is just one time-symmetrization instance and it obviously involves the same number of oracle queries of the original quantum algorithm.

As one can see -- table (\ref{disr}) -- the computational
complexity of Grover's problem reduces in the time-symmetrized quantum algorithm with respect to the problem solver Alice. Alice's problem is now locating the ball hidden in the pair
of drawers
$\left \{01 ,11\right \}$
-- check the input and output states of
$\hat{\mathbf{U}}_{1 ,2}$
in that table. Let us call the new problem \textit{the reduced problem}.

In equivalent terms, table (\ref{disr}) tells us that Alice, immediately after the initial
measurement, knows that the ball is in the pair of drawers
$\left \{01 ,11\right \}$. In fact, in the description of the quantum algorithm with respect to her, the state of register
$B$
represents her state of knowledge of the number
of the drawer with the ball. In the input state of her problem-solving action, this changes from knowledge that the ball is in one of four drawers (i.e. from
complete ignorance of the number of the drawer with the ball), into knowledge that the ball is in one of two drawers --
compare
the states in the top-left and bottom-left corners of table (\ref{syal}). This is because Alice is shielded from the half information
about the number of the drawer with the ball coming to her from the initial measurement, not from
the half coming to her back in time from the final measurement.

Summing up, all is as if Alice knew in advance half of the information about the solution of the problem she will produce and read in the
future and could use this knowledge to reduce the computational complexity of the problem to be
solved. Below, we show that this allows to compute the number of oracle queries needed to solve
Grover's problem in an optimal quantum way (of course without knowing the quantum algorithm).

 The reduced problem, as any other problem,
can always be solved quantumly with the number of oracle queries required by a (reversible) classical
algorithm -- i.e. logically required\protect\footnote{ What face each other here are not quantum and classical physics but
quantum physics and classical logic.
}. The question then becomes whether it could be solved with even fewer
queries, namely itself with a quantum speedup. Now, under the reasonable assumption that the quantum
speedup is essentially related to the reduction, under time-symmetrization, of the computational
complexity of the problem to be solved, the answer must be negative. In fact, the quantum algorithm
that solves the reduced problem has been time-symmetrized already. Further time-symmetrizing it
would leave it unaltered without further reducing the computational complexity of the
problem\protect\footnote{ We have seen that Bob and Alice can effectively perform the partial
measurements of the instance we are dealing with. The time-symmetrization in this case consists in propagating
forward in time the selection performed by Bob's measurement and backwards in time that performed by
Alice's measurement. Doing this a second time changes nothing.
}.

Summing up, it turns out that the number of
queries needed to solve Grover's problem in an optimal quantum way is that needed to solve the
corresponding reduced problem in an optimal logical (classical) way. In the present four drawer
instance of Grover's problem, this is only one query. In fact, the reduced problem, i.e. locating a ball
hidden in a pair of drawers, can of course be classically solved by opening either drawer.
Note that we have found the number of oracle queries needed to solve Grover's problem in
an optimal quantum way without knowing Grover algorithm, namely
$\hat{\mathbf{U}}_{1 ,2}$.

More in general, we should consider the
generic number of drawers
$N =2^{n}$, where
$n$
is the number of binary digits (bits) in the drawer number. Also note that, in the variant provided
by Long
$\left [23\right ]$, Grover algorithm gives the solution with certainty for any value of
$N$. Everything we said for the four drawers still holds. 

In order that the initial and final measurements evenly contribute to the selection of the pair of measurement outcomes, it suffices to evenly share between the two measurements
the selection of the
$n$
bits that specify the number of the drawer with the ball. This means ascribing to the final
measurement the selection of
$n/2$
bits.              Since Alice knows these bits in advance, the reduced problem is
locating the ball in
$2^{n/2}$
drawers. It can be solved logically (classically) with
$\ensuremath{\operatorname*{O}} \left (2^{n/2}\right )$
queries.

Summarizing, the time-symmetrization
formalism foresees that the number of queries needed to solve
Grover's problem in an optimal quantum way is that needed to solve the corresponding reduced
problem in an optimal logical (classical) way, namely
$\ensuremath{\operatorname*{O}} \left (2^{n/2}\right )$
queries. Also in the quantum case we must make reference to an optimal
quantum algorithm since a non optimal one could take any higher number of queries. That Grover's
problem can be solved quantumly with any number of queries above that required by the optimal
Grover/Long algorithm has been shown in
$\left [23 ,24\right ]$.

All the above is in agreement with the fact
that Grover algorithm (more precisely Grover/Long algorithm) is demonstrably optimal
$\left [ 23 -25\right ]$
and employs
$\ensuremath{\operatorname*{O}} \left (2^{n/2}\right )$
queries $\left [5\right ]$.

 Let us eventually note that the present explanation of the quantum speedup ascribes it to a quantum causal loop: the final measurement of the solution changes back in time the initial state of
knowledge of the solution on the part of Alice and this in turn allows Alice to produce the solution with fewer oracle queries, what closes the loop.

\subsubsection{ From Grover's problem to any oracle problem
}
The method to compute the number of oracle queries needed to solve Grover's problem in an
optimal quantum way can immediately be extended to any oracle problem, as follows.

In any quantum algorithm relativized to Alice, we have:

(i) an initial state which is the superposition of all the possible problem-settings tensor product the blank blackboard that will eventually contain the solution of the problem and 

(ii) a state immediately before the final Alice's measurement of the solution, which is the superposition of tensor product, each a problem-setting multiplying the corresponding solution.

In other words, provided that we change the size of the quantum registers, the diagrams developed for Grover algorithm hold for the process of solving any oracle problem.

The only possible difference with respect to Grover's case is that the solution is a non-invertible function of the problem-setting. 

This difference is eliminated by thinking that Alice, at the end, measures both the content of register $A$, which contains the solution, and that of register $B$, which contains the problem-setting. This is legitimate since this Alice's measurement of the content of $B$ occurs after the unitary part of her problem solving action, namely when the reason for hiding from her the problem-setting has fallen. 

The time-symmetrization of this process requires that, in each time-symmetrization instance, the two measurements of the content of register $B$ -- the initial measurement made by Bob and the final one made by Alice\protect\footnote{
Alice's measurement of the solution is redundant with her measurement of the problem-setting and can be ignored.
} -- reduce to two partial measurements that evenly and non-redundantly contribute to the selection of the problem-setting and the corresponding solution. Eventually, we should take the quantum superposition of all the time-symmetrization instances.

Let us further introduce a simplification. Since the reduced density operator of register $B$ remains unaltered throughout the unitary part of Alice's action, any final measurement of a content of $B$ on the part of Alice can be anticipated to the time of the initial measurement. The requirement of evenly sharing
between the initial Bob's measurement and the final
Alice's measurement the selection of the pair of correlated measurement outcomes can be replaced by that of splitting the initial measurement of the content of register $B$ into two partial measurements that
evenly and non redundantly contribute to selecting the problem-setting and the corresponding solution. In each time-symmetrization instance, Alice knows in advance the information selected by either partial measurement. Taking this into account yields the following
\textit{advanced knowledge rule} that holds for any oracle problem:

\medskip

\textit{We should split the initial
measurement of the problem-setting}
\textit{ in a state of maximal indetermination of it into two partial measurements
that evenly and non-redundantly contribute to the selection of the problem-setting}
\textit{ and the corresponding solution}\textit{. Alice knows in advance the
information}\textit{ acquired by either partial measurement. The number of queries needed
to solve the oracle problem in an optimal quantum way is that needed to solve it in an optimal
logical (classical) way given the advanced knowledge in question.}

\medskip

Let us consider for example the problem solved by Deutsch\&Jozsa algorithm $\left [26\right ]$. The set of functions computed by the black box is all the constant and \textit{balanced} functions $f_{\mathbf{b}} :\left \{0 ,1\right \}^{n} \rightarrow \left \{0 ,1\right \}$ -- balanced functions have an even number of zeros and ones. The following table gives four of the eight functions for $n =2$.

\begin{equation*}\begin{array}{cccccc}\mathbf{a} & f_{0000}\left (\mathbf{a}\right ) & f_{1111}\left (\mathbf{a}\right ) & f_{0011}\left (\mathbf{a}\right ) & f_{1100}\left (\mathbf{a}\right ) &  . . . \\
00 & 0 & 1 & 0 & 1 &  . . . \\
01 & 0 & 1 & 0 & 1 &  . . . \\
10 & 0 & 1 & 1 & 0 &  . . . \\
11 & 0 & 1 & 1 & 0 &  . . .\end{array}
\end{equation*}

The first column from the left contains the argument of the function, the second and third the corresponding values of the two constant functions, the fourth those of a balanced function, etc. Note that we have chosen as problem-setting (and function suffix) $\mathbf{b}$ the table of the function -- the sequence of function values for increasing values of the argument. 

Bob selects one of these functions -- a valuation of $\mathbf{b}$ -- at random, Alice is to find whether $f_{\mathbf{b}}$ is constant or balanced by performing oracle queries. In the classical and worst case, an $\exp \left (n\right )$ number of queries is needed, in the quantum case just one.

We apply the advanced knowledge rule to this algorithm. The outcome of the initial Bob's measurement is the problem-setting $\mathbf{b}$. We should split this measurement into two partial measurements that evenly and non-redundantly contribute to the selection of $\mathbf{b}$  and the corresponding solution -- Alice knows in advance the information acquired by either partial measurement.

 As one can see, this implies that these two partial measurements measure two halves of the bit-string $\mathbf{b}$ that contain either all zeros or all ones -- we call such halves \textit{good half tables}. For example, let $\mathbf{b} =0011$ -- the table of the corresponding function is the fourth column of the above table. The two good half tables are $f_{0011}\left (00\right ) =0 ,f_{0011}\left (01\right ) =0$ and $f_{0011}\left (10\right ) =1 ,f_{0011}\left (11\right ) =1$. If either half table contained both zeros and ones, also the other would. This would determine the solution (that the function is balanced) in a redundant way (and, of course, Alice would know the solution of the problem without doing anything). The advanced knowledge rule would be violated.

Then, Alice knows in advance a good half table -- e. g. $f_{\mathbf{b}}\left (00\right ) =0 ,f_{\mathbf{b}}\left (01\right ) =0$. We have omitted the value of the function suffix because of course she only knows the result of measuring the content of the left two digits of the problem-setting $0011$ contained in register $B$, she does not know the suffix of the function. Once she knows in advance a good half table, classically, performing a single oracle query for a value of the argument outside that half table (here e. g. for $\mathbf{a} =10$) allows her to ascertain whether the function is constant or balanced, in agreement with what Deutsch\&Jozsa algorithm does.

Let us now go to Simon $\left [27\right ]$ and the quantum part of Shor $\left [28\right ]$ factorization algorithm. Both these algorithms concern finding the period of a particular set of periodic functions. Classically, this requires a number of oracle queries exponential in problem size, quantumly the number becomes polynomial -- there is thus an exponential speedup. Applying the advanced knowledge rule to these algorithms yields what follows.

One can choose as problem-setting $\mathbf{b}$ the table of two consecutive periods of the function. Naturally, the period of the function is identified by two consecutive values of the argument of the function such that the corresponding values of the function are the same. One can see that the notion of good half table still applies and that a pair of good half tables are the tables of the two periods.

Since Alice knows in advance a good half table (i.e. the table of a period of the function) performing an oracle query for a value of the argument immediately outside that half table necessarily yields a repeated value of the function, what allows her to find the period of the function.  It should be noted that, according to the advanced knowledge rule, the algorithms of Simon and Shor would be suboptimal, as they require a number of oracle queries polynomial in problem size instead of a single query -- see $\left [4\right ]$ for further detail.

Summing up, the advanced knowledge rule exactly accounts for the speedups of Grover and Deutsch\&Jozsa algorithms, which are demonstrably optimal, and in good approximation for those of Simon and Shor algorithms -- more in
general for those of the Abelian hidden subgroup
$\left [29\right ]$. As is well known, these are the major quantum algorithms discovered until now and
cover both the quadratic and exponential speed ups. 

We note that in all cases the speedup is ensured by a quantum causal loop. The final measurement of the problem-setting on the part of the problem solver changes backwards in time her initial state of complete ignorance of it (and of the corresponding solution) into knowledge of half of the information that specifies it. This knowledge can be used by the problem solver to reach the solution with fewer oracle queries, what closes the loop.

\section{ Time-symmetrizing the ordinary quantum description of nonlocality
}

We go now quantum nonlocality. We assume that two photons $B$ and $A$ at time $t_{0}$ have been generated by parametric down-conversion in a state of spatial contiguity and in the polarization state: \begin{equation}\frac{1}{\sqrt{2}}\left (\vert 0\rangle _{B}\left \vert 1\right \rangle _{A} -\vert 1\rangle _{B}\left \vert 0\right \rangle _{A}\right ) , \label{sym}
\end{equation}  $0/1$
are horizontal/vertical polarization. Then, in the time interval
$\left (t_{0} ,t_{1}\right )$, the two photons get spatially separate in a unitary way. The polarization of photon $B$ is measured at time $t_{1}$, that of photon $A$ is measured in the same basis at time $t_{2}$. There is one-to-one correlation between the two measurement outcomes and, correspondingly, a unitary transformation between them. This transformation can go  from $t_{1}$ to $t_{2}$ either directly or via $t_{0}$, first backwards in time from $t_{1}$ to $t_{0}$ and then forward in time from $t_{0}$ to $t_{2}$. We should keep in mind that the unitary transformation that
connects the two measurement outcomes via
$t_{0}$
is local and thus physical (of course, at the condition that we consider the present form of quantum retrocausality physical too). The one
that goes directly from
$t_{1}$
to $t_{2}$, implying the spooky action at a distance, is certainly unphysical.

To explain quantum nonlocality by means of the time-symmetrization (TS)  formalism, we should time-symmetrize the reversible quantum process that goes from $t_{1}$ to $t_{2}$ via $t_{0}$. We should evenly share
between the two measurements the selection of the single bit that
specifies the pair of correlated outcomes among the two possible pairs $01$ and $10$. To this end, we represent the polarization state (\ref{sym}) in the redundant way:
\begin{equation}\frac{1}{2}\left (\vert 00 \rangle _{B}\vert  00 \rangle _{A} +\vert 01 \rangle _{B}\vert  01 \rangle _{A} +\vert 10 \rangle _{B}\vert  10 \rangle _{A} +\vert 11 \rangle _{B}\vert  11 \rangle _{A}\right ) . \label{exp}
\end{equation}

One can see that, if we take the XOR (exclusive
OR) between the two bits of each register, we obtain the state
$\frac{1}{\sqrt{2}}\left (\vert 0\rangle _{B}\left \vert 0\right \rangle _{A} +\vert 1\rangle _{B}\left \vert 1\right \rangle _{A}\right )$. If, in (\ref{exp}),
we changed the two central plus signs into minus signs, we would have obtained
$\frac{1}{\sqrt{2}}\left (\vert 0\rangle _{B}\left \vert 0\right \rangle _{A} -\vert 1\rangle _{B}\left \vert 1\right \rangle _{A}\right )$, which can trivially be brought to state (\ref{sym}).
However, we will do without these changes that are irrelevant here since we never rotate any
measurement basis. Now the problem is the same as that of the four drawer instance of Grover algorithm:
evenly (and non-redundantly) sharing between the measurements of
$\hat{B}$
and
$\hat{A}$
the selection of the two bits that specify the pair of correlated outcomes.

\subsection{ Costa de Beauregard's explanation
}
 First, we provide the ordinary (usual) quantum description of nonlocality (we always ignore normalization):

\begin{equation}\begin{array}{ccc}\text{time }t_{1}\text{, }\text{}\text{meas. of}\;\hat{B} & t_{1} \rightarrow t_{2} & \text{time }t_{2}\text{}\text{}\text{}\text{, meas. of }\text{}\hat{A} \\
\, & \, & \, \\
\vert 00 \rangle _{B}\vert  00 \rangle _{A} +\vert 01 \rangle _{B}\vert  01 \rangle _{A} +\vert 10 \rangle _{B}\vert  10 \rangle _{A} +\vert 11 \rangle _{B}\vert  11 \rangle _{A} & \, & \, \\
\Downarrow  & \, & \, \\
\vert 01 \rangle _{B}\vert  01 \rangle _{A} &  \Rightarrow \hat{\mathbf{U}}_{1 ,2} \Rightarrow  & \vert 01 \rangle _{B}\vert  01 \rangle _{A}\end{array} \label{usuale}
\end{equation}
At time
$t_{1}$, the measurement of$\;\hat{B}$
(the content of register
$B$) in the maximally entangled state in the top-left of the table selects at random, say, the
eigenvalue
$b =01$,
thus projecting the state of the two registers on
$\vert 01 \rangle _{B}\vert  01 \rangle _{A}$
(vertical arrow). The fact that this measurement instantly changes also the state of the
space separate register
$A$
is of course the ``spooky action at a distance''. The successive unitary transformation
$\hat{\mathbf{U}}_{1 ,2}$, between $t_{1}$ and $t_{2}$, is the identity for the polarization state of the two photons. By
$\hat{\mathbf{U}}_{1 ,2}$, the measurement outcome
$\vert 01 \rangle _{B}\vert  01 \rangle _{A}$ directly (not via
$t_{0}$) propagates to immediately before the measurement of
$\text{}\hat{A}$
(the content of register
$A$) at time $t_{2}$ -- horizontal arrows. Then this latter measurement deterministically selects the
correlated eigenvalue
$a =01$.

The following table provides the quantum
process corresponding to Costa de Beauregard's explanation of quantum nonlocality
$\left [9 ,30\right ]$.
Formally, it suffices to replace, in table (\ref{usuale}),
$\hat{\mathbf{U}}_{1 ,2}$ by $\hat{\mathbf{U}}_{1 ,0 ,2}$, where
$\hat{\mathbf{U}}_{1 ,0 ,2}$ is the unitary transformation that goes from the first to the second measurement outcome, one-to-one correlated with it, via $t_{0}$. Of course we have  the mathematical identity
$\hat{\mathbf{U}}_{1 ,2} =\hat{\mathbf{U}}_{1 ,0 ,2}$. Table (\ref{usuale}) becomes:

\begin{equation}\begin{array}{ccc}\text{time }t_{1}\text{, }\text{}\text{meas. of}\;\hat{B} & t_{1} \rightarrow t_{0} \rightarrow t_{2} & \text{time }t_{2}\text{, mes. of }\text{}\hat{A} \\
\, & \, & \, \\
\vert 00 \rangle _{B}\vert  00 \rangle _{A} +\vert 01 \rangle _{B}\vert  01 \rangle _{A} +\vert 10 \rangle _{B}\vert  10 \rangle _{A} +\vert 11 \rangle _{B}\vert  11 \rangle _{A} & \, & \, \\
\Downarrow  & \, & \, \\
\vert 01 \rangle _{B}\vert  01 \rangle _{A} &  \Rightarrow \hat{\mathbf{U}}_{1 ,0 ,2} \Rightarrow  & \vert 01 \rangle _{B}\vert  01 \rangle _{A}\end{array} \label{costad}
\end{equation}

This time the outcome of measuring$\;\hat{B}$
at time
$t_{1}$
unitarily propagates first backwards in time to
$t_{0}$ then forward in time to
$t_{2}$, becoming the state immediately before and after the measurement of
$\text{}\hat{A}$.

The point is that table (\ref{costad}) hosts Costa de Beauregard's explanation of quantum nonlocality. One
should assume that, in causal order, the measurement of$\;\hat{B}$
at time $t_{1}$ only projects the reduced density operator of register
$B$
on
$\left \vert 01\right \rangle _{B}\left \langle 01\right \vert _{B}$, leaving that of register
$A$ unaltered. Then this projection propagates backwards in time from
$t_{1}$
to
$t_{0}$. At time $t_{0}$, it locally projects the original maximally entangled state on the sharp state
$\vert 01 \rangle _{B}\vert  01 \rangle _{A}$. Eventually the latter state propagates forward in time from
$t_{0}$
to
$t_{2}$ where it becomes the state immediately before and after the final measurement. 

Note that, under this interpretation, the fact that the measurement of\textbf{$\;\hat{B}$} at time
$t_{1}$
instantly changes also the state of the space separate register
$A$
is, so to speak, a coincidence. It does instantly change the state of register
$A$, but by two causally successive local propagations, one backwards in time from
$t_{1}$
to
$t_{0}$ and the other forward in time from
$t_{0}$
to
$t_{1}$. Of course, these two propagations take the same amount of time, but one with a negative and the other with a positive sign, so that the overall time taken is zero. This is what emulates action at a distance.

 Costa de Beauregard's
explanation received little attention by the majority of physicists
$\left [30\right ]$. It is difficult to say why because one should explain silence. Likely, the notion that
causality can go backwards in time, even along the objectively ideal reversible processes of quantum
mechanics, is hardly accepted. In the next subsection, we will rebuild a similar explanation by
means of the time-symmetrization formalism, thus replacing usual retrocausaliy by its present unobservable form.

\subsection{ Application of the time-symmetrization formalism
}
 First, we apply the time-symmetrization (TS) formalism
to the usual description
of quantum nonlocality of table (\ref{usuale}), where
the outcome of the first measurement directly propagates, by
$\hat{\mathbf{U}}_{1 ,2}$, to the second
measurement. Note that we are dealing with a description with respect to any observer. We do not
have to hide anything from anyone here.
We consider the time-symmetrization instance where the sorted out value of the two
correlated measurement outcomes is
$01$, the measurement of
$\hat{B}$
at time
$t_{1}$
reduces to that of
$\hat{B}_{l}$
(the left bit
of the number contained in register
$B$) and the measurement of
$\hat{A}$
at time
$t_{2}$
to that of
$\hat{A_{r}}$
(the right bit
of the number in register
$A$).  The corresponding zigzag diagram is:

\begin{equation}\begin{array}{ccc}\text{time }t_{1}\text{, }\text{}\text{meas. of 
$\hat{B}_{l}$} & t_{1} \rightleftarrows t_{2} & \text{time }t_{2}\text{, }\text{}\text{meas. of}\text{$\hat{A_{r}}$} \\
\, & \, & \, \\
\vert 00 \rangle _{B}\vert  00 \rangle _{A} +\vert 01 \rangle _{B}\vert  01 \rangle _{A} +\vert 10 \rangle _{B}\vert  10 \rangle _{A} +\vert 11 \rangle _{B}\vert  11 \rangle _{A} & \, & \, \\
\Downarrow  & \, & \, \\
\vert 00 \rangle _{B}\vert  00 \rangle _{A} +\vert 01 \rangle _{B}\vert  01 \rangle _{A} &  \Rightarrow \hat{\mathbf{U}}_{1 ,2} \Rightarrow  & \vert 00 \rangle _{B}\vert  00 \rangle _{A} +\vert 01 \rangle _{B}\vert  01 \rangle _{A} \\
\, & \, & \Downarrow  \\
\vert 01 \rangle _{B}\vert  01 \rangle _{A} &  \Leftarrow \hat{\mathbf{U}}_{1 ,2}^{\dag } \Leftarrow  & \vert 01 \rangle _{B}\vert  01 \rangle _{A}\end{array} \label{precu}
\end{equation}
The measurement of
$\hat{B}_{l}$
at time
$t_{1}$
projects the initial state superposition (top-left of the table) on the superposition of
the terms beginning with
$0$
(vertical arrow on the left of the table). The latter superposition, by
$\hat{\mathbf{U}}_{1 ,2}$, propagates to
$t_{2}$
(right looking horizontal arrows), where the measurement of
$\hat{A_{r}}$
projects it on the sharp state
$\vert 01 \rangle _{B}\vert  01 \rangle _{A}$
ending with
$1$
(vertical arrow on the right of the table). This measurement outcome, by
$\hat{\mathbf{U}}_{1 ,2}^{\dag }$, propagates to
$t_{1}$
(left looking horizontal arrows). The propagation in question, which inherits both
selections, is both a time-symmetrization instance and the superposition of all instances for
measurement outcome
$01$
-- the bottom lines of all these instances being identical to each other.
They are also identical to the ordinary quantum description -- bottom line of table (\ref{usuale}) -- since ``$ \Leftarrow \hat{\mathbf{U}}_{1 ,2}^{\dag } \Leftarrow $'' is equivalent to ``$ \Rightarrow \hat{\mathbf{U}}_{1 ,2} \Rightarrow $''.

Now we apply the TS formalism
to the description
of the quantum process that goes from
$t_{1}$
to
$t_{2}$
via
$t_{0}$. It suffices to replace,  in table (\ref{precu}),
$\hat{\mathbf{U}}_{1 ,2}$
by the mathematically equivalent
$\hat{\mathbf{U}}_{1 ,0 ,2}$. This yields:

\begin{equation}\begin{array}{ccc}\text{time }t_{1}\text{, }\text{}\text{meas. of 
$\hat{B}_{l}$} & t_{1} \rightleftarrows t_{0} \rightleftarrows t_{2} & \text{time }t_{2}\text{, }\text{}\text{meas. of}\text{$\hat{A_{r}}$} \\
\, & \, & \, \\
\vert 00 \rangle _{B}\vert  00 \rangle _{A} +\vert 01 \rangle _{B}\vert  01 \rangle _{A} +\vert 10 \rangle _{B}\vert  10 \rangle _{A} +\vert 11 \rangle _{B}\vert  11 \rangle _{A} & \, & \, \\
\Downarrow  & \, & \, \\
\vert 00 \rangle _{B}\vert  00 \rangle _{A} +\vert 01 \rangle _{B}\vert  01 \rangle _{A} &  \Rightarrow \hat{\mathbf{U}}_{1 ,0 ,2} \Rightarrow  & \vert 00 \rangle _{B}\vert  00 \rangle _{A} +\vert 01 \rangle _{B}\vert  01 \rangle _{A} \\
\, & \, & \Downarrow  \\
\vert 01 \rangle _{B}\vert  01 \rangle _{A} &  \Leftarrow \hat{\mathbf{U}}_{1 ,0 ,2}^{\dag } \Leftarrow  & \vert 01 \rangle _{B}\vert  01 \rangle _{A}\end{array} \label{tsm}
\end{equation}
Of course, also in this case the bottom lines of the time-symmetrization instances are
all identical to each other, and to that of the ordinary quantum description of table (\ref{costad}) -- we can replace ``$ \Leftarrow \hat{\mathbf{U}}_{1 ,0 ,2}^{\dag } \Leftarrow $'' by ``$ \Rightarrow \hat{\mathbf{U}}_{1 ,0 ,2} \Rightarrow $''. The important point is that the projections of both reduced density operators, that of register
$B$
after the measurement of
$\hat{B}_{l}$
and that of  register
$A$
after the measurement of
$\hat{A_{r}}$, in their propagation toward the other measurement go via
$t_{0}$. At time
$t_{0}$
they locally project the state of the entire quantum system. Also the present
zigzag diagram hosts the local explanation of spatial nonlocality.

 By  the way, one might wonder how the projection of the reduced density operator of either register "knows" that it must propagate to the other measurement through $t_{0}$.  Our answer would be: it propagates in all the possible ways along the unitary evolutions that connect the two measurement outcomes.

Eventually, let us note that what explains
spatial nonlocality are a couple of quantum causal loops, one for each measurement. The two
measurements, respectively at times
$t_{1}$
and
$t_{2}$, change back in time the state of the two registers at time
$t_{0}$. This in turn causes the correlation between the two future measurement outcomes.

 Let us compare now the explanation of nonlocality provided by the TS formalism and
that of Costa de Beauregard.

The latter explanation assumes that the entire outcome of  measuring $\hat{B}$
at time
$t_{1}$(which only changes the reduced density operator of register $B$) goes backwards in time to the time
$t_{0}$
the two subsystems were not spatially separate. This: (i) violates the present requirement that the description of quantum correlation does not present a preferred time-direction of causality, (ii) changes a past state as described by the ordinary quantum description and (iii) can be considered an
arbitrary assumption in the sense that it is only justified by the explanation
it provides.

 The TS formalism replaces
the ordinary retrocausality of Costa de Beauregard's explanation of table (\ref{costad}) by the unobservable retrocausality of
table (\ref{tsm}). The latter comes from time-symmetrizing the ordinary quantum description
of the reversible process that physically connects, via $t_{0}$, the two measurement outcomes. In present
assumptions, more than being legitimate, this time-symmetrization is mandatory to complete the quantum description.
Furthermore, the two not yet space-separate subsystems are told about how to behave in
the future measurements by hidden variables: the unobservable time-directions of the causations that physically ensure the correlation between the two outcomes. This unobservable way
of sending information backwards in time that does not change the observable past should be acceptable.

\section{
Discussion}

In this section we discuss the unconventional / unorthodox points of the present work. They can be
summarized as follows:

(i) Opening to retrocausality.

(ii) Resuming the apparently forgotten
problem of the incompleteness of the quantum description raised by Einstein, Podolsky and
Rosen in their famous EPR paper.

 (iii) Tackling the same problem in the quantum speedup. Its ordinary quantum description is a reversible process between two one-to-one correlated measurement outcomes exactly as in the case of quantum nonlocality and thus must suffer from the same incompleteness.

 (iv) Deriving a technical result -- the rule
for computing the number of oracle queries needed to solve an oracle problem in an optimal quantum
way -- in an axiomatic way from physical principles.

 (v) Showing that the quantum capability of
performing tasks that cannot be performed classically essentially relies on unobservable causal loops.

 Any of the above points can be puzzling. We
defend them as follows:

\medskip

 Point (i). The fact that retrocausality can change the past, in the macroscopic world
gives rise to well-known paradoxes. However, the present form of quantum retrocausality has a
justification and a remedy. The justification is that retrocausality is generated by a mandatory operation: completing the ordinary quantum description by time-symmetrizing it. The remedy is that it resides in each unobservable
time-symmetrization instance and vanishes in their quantum superposition, namely in the ordinary
quantum description. In other words, the quantum form of retrocausality generated by the time-symmetrization formalism is unobservable and does
not change the past as described by the ordinary quantum description. It should therefore be immune to the paradoxes that plague the usual form.

Points (ii) - (iii).  In hindsight, it is clear that the ordinary description of quantum correlation -- hence of nonlocality and the speedup -- is incomplete. Indeed, it cannot describe the causal loops on which both are based. By completing it, it does.

\medskip

Point (iv). The number of oracle queries
needed to solve an oracle problem in an optimal quantum way has been axiomatically derived from the
principle that the quantum description of a process which is reversible from the initial to the final measurement outcome must be symmetric in time -- must not entail a preferred time-direction of causality. This derivation solves in a synthetic way the so called quantum \textit{query complexity problem}, which is central to quantum computer science and still unsolved in the general case
$\left [7 ,8\right ]$. The synthetic character of this derivation contrasts with the analytical (entirely mathematical)
character of the derivations typical of the latter discipline. Our defense is that the two kinds of derivation should be
equivalent if quantum mechanics is consistent.

In favor of the synthetic approach, we would invoke Grover's authority. In 2001
$\left [31\right ]$, he called for a synthetic demonstration of the optimality of his quantum search
algorithm. He wrote: \textit{It has been proved that the
quantum search algorithm cannot be improved at all, i.e. any quantum mechanical algorithm will need
at least }
$\ensuremath{\operatorname*{O}}\sqrt{N}$
\textit{ steps to carry out an exhaustive search of }
$N$
\textit{ items. Why is it not possible to search in fewer than }
$\ensuremath{\operatorname*{O}}\sqrt{N}$
\textit{ steps? The argument used to prove this are very subtle and mathematical. What
is lacking is a simple and convincing two line argument that shows why one would expect this to be
the case}.

 Our ``two line'' argument would be that time-symmetrizing the ordinary quantum description of the reversible process of setting and solving Grover's problem reduces its quantum computational complexity from
$\ensuremath{\operatorname*{O}}\left (N\right )$
to
$\ensuremath{\operatorname*{O}}\sqrt{N}$. Further time-symmetrizing the already time-symmetrized description would leave it unaltered -- would not further reduce the problem complexity.

\medskip

Point (v). We have shown that, in the two
known cases of the quantum speedup and quantum nonlocality, the quantum ability to perform tasks that cannot be performed classically relies on
causal loops. Accepting the idea that non-relativistic quantum mechanics can host causal loops might
be difficult. However, the fact that such loops reside in the unobservable time-symmetrization
instances and vanish in their quantum superposition, namely in the ordinary quantum description, should make it easier.

\section{ Summary and conclusion
}
 As the present work is rather unconventional, it may be useful to summarize it before concluding.

\subsection{ Summary
}
We have been dealing with the quantum description of the reversible processes between two one-to-one
correlated measurement outcomes that characterize both the quantum speedup and quantum nonlocality. We have argued that the ordinary quantum description of these processes mathematically describes the correlation between the two outcomes but leaves the causal structure that physically ensures it free.

We have seen that, consequently, the selection of the information that specifies the sorted out pair of outcomes among all the possible pairs can share in any way between the initial and final measurements. 

In present assumptions, the ways where this sharing is uneven violate the time-symmetry required of the description of a time-reversible process: the fact it should not entail a preferred time-direction of causality. This says that the ordinary quantum description, which allows such violations, is incomplete and is completed by time-symmetrizing it. To this end, the selection of the information that specifies the pair of correlated measurement outcomes should be evenly (and non-redundantly) shared between the initial and final measurements. Since this can be done in a plurality of ways, we should take their quantum superposition.

This time-symmetrization of the ordinary quantum description leaves it unaltered, as
it should be, but at the same time shows that it is a quantum superposition of unobservable
time-symmetrization instances whose causal structure is completely defined. Each instance is a causal loop. Causality goes from the initial to the final measurement outcome and then back from the latter to the initial measurement outcome.

These instances are unobservable since they vanish in the superposition of all instances. In fact the latter is the ordinary quantum description, which leaves the causal structure that physically ensures the correlation between the two measurement outcomes free. Being completed descriptions of the quantum process, they describe it in a more complete way than the ordinary quantum description. In particular, they immediately explain the quantum speedup and quantum nonlocality.

In the case of the quantum speedup, the final measurement of the problem-setting and the corresponding solution on the part of the problem-solver changes backwards in time the initial measurement outcome. The latter represents both the problem to be solved and the knowledge of the problem-setting and the corresponding solution on the part of the problem solver.  The computational complexity of the problem to be solved reduces. Correspondingly, the problem solver's complete
ignorance of the problem-setting and the corresponding solution changes into knowledge of half of the information that specifies them. The
causal loop becomes the fact that all is as if the problem solver knew in advance half of the information about
the problem-setting and the corresponding solution she will read in the future and could use this knowledge to produce the solution with
fewer oracle queries.

 In the case of quantum nonlocality, the measurement
on either subsystem, when the two are space-separate, retrocausally and locally changes the state of both subsystems when they were not space-separate. The causal loop implicit in the time-symmetrization formalism originates two causal loops, one for each measurement. Together, they locally cause
the correlation between the two measurement outcomes.

\subsection{ Conclusion
}
In conclusion, the key finding of the present approach is that the ordinary quantum description of the processes between two one-to-one correlated measurement outcomes, which leaves the causal structure that physically ensures the correlation free, is the quantum superposition of unobservable casual loops. Such causal loops, which of course vanish in the ordinary quantum description, describe the quantum process in a more complete way. In particular, they immediately explain the quantum speedup and quantum nonlocality.

Future work, at the fundamental level, could be to look  for causal loops
in any task that is quantumly possible and classically impossible and for the possibility of
extending the present time-symmetrization formalism beyond one-to-one quantum correlation. Of course, if the probability of finding the solution is not exactly one, there can still be a quantum speedup and a causal loop that ensures it. At the technical level, the advanced knowledge rule could be used in the search for new quantum algorithms and to arrange oracle problems into quantum complexity classes. One could also
see whether the present axiomatic derivation of the rule for finding the number of queries needed to
solve an oracle problem in an optimal quantum way might suggest a corresponding analytic derivation.
About the possibility of cross fertilization between the foundations of physics and quantum computer
science, see
$\left [32\right ]$.

We hope that the present work will help to clear through customs the notion of retrocausality in physics, in its present form of unobservable quantum causal loops that vanish in the ordinary quantum description and thus cannot change the past as described by it. This notion could open up to physics a territory that was previously off limits.

\section*{ Acknowledgments
}
 Thanks are due to Eliahu Cohen, Artur Ekert, Avshalom Elitzur, David
Finkelstein, and Ken Wharton for useful suggestions and to Daniel Sheehan for organizing the San Diego
AAAS-Pacific Division Congresses on Retrocausalty, a forum of discussion that has been important for
the development of the present approach.

\section*{ References
}

$\left [1\right ]$
Castagnoli, G. and Finkelstein D. R.. Theory of the quantum speedup. \textit{Proc. Roy. Soc. Lon.}, 457 (2001), 1799-1807.

$\left [2\right ]$
Castagnoli, G.. The quantum correlation between the selection of
the problem and that of the solution sheds light on the mechanism of the quantum speed up.
\textit{Phys. Rev. A} 82 (2010), 052334.

$\left [3\right ]$
Castagnoli, G.. Completing the Physical Representation of Quantum
Algorithms Provides a Quantitative Explanation of Their Computational Speedup. Found. Phys. 48
(2018), 333-354.

$\left [4\right ]$
Castagnoli, G.,  Cohen, E., Ekert, A. K., and  Elitzur, A. C.. A Relational
Time-Symmetric Framework for Analyzing the Quantum Computational Speedup. \textit{Found Phys.}, 49, 10 (2019),            1200-1230.

 $\left [5\right ]$
Grover, L. K.. A fast quantum mechanical algorithm for database
search. \textit{Proc. 28th Annual ACM Symposium on the Theory
of Computing}. ACM press New York (1996), 212-219.

$\left [6\right ]$
Einstein, A., Podolsky, B., and Rosen, N.. Can Quantum-Mechanical Description of
Physical Reality Be Considered Complete? \textit{Phys.
Rev.} 47, (1935),
777-780.

$\left [7\right ]$
Aaronson S., Ambainis, A., Iraids, J., Kokainis, M., Smotrovs, J.. Polynomials, Quantum
Query Complexity, and Grothendieck's Inequality. \textit{CCC'16: Proceedings of the 31st Conference
on Computation and Complexity}, May 2016 and arXiv:1511.08682v3 [quant-ph] (2016)

$\left [8\right ]$
Ambainis, A.. Understanding Quantum Algorithms via Query Complexity.
arXiv:1712.06349v1 [quant-ph] (2017)

$\left [9\right ]$
Costa de Beauregard, O.. Mechanique quantique.\textit{ Compte Rendus Academie des Siences}, 236 (1953),
1632-34.

$\left [10\right ]$
Wheeler, J. A., Feynman, R. P.. Interaction with the Absorber as the Mechanism of
Radiation. \textit{Rev. Mod. Phys.}, vol 17, 2-3 (1945), 157-161.

$\left [11\right ]$ Aharonov, Y., Bergman, P. G., and Lebowitz, J. L.. Time Symmetry
in the Quantum Process of Measurement. \textit{Phys. Rev.} 134 (1964),
1410-1416.

$\left [12\right ]$
Dolev, S. and Elitzur, A. C.. Non-sequential behavior of the wave
function. arXiv:quant-ph/0102109 v1 (2001).

$\left [13\right ]$
Aharonov, Y. and Vaidman, L.. The Two-State Vector Formalism: An
Updated Review. \textit{Lect. Notes Phys.} 734 (2008), 399-447.

$\left [14\right ]$
Aharonov, Y., Cohen, E., and Elitzur, A. C.. Can a future choice
affect a past measurement outcome? \textit{Ann.
Phys.} 355 (2015), 258-268.

$\left [15\right ]$
Aharonov, Y., Cohen E., and Tollaksen, J.. Completely top-down
hierarchical structure in quantum mechanics. \textit{Proc. Natl. Acad. Sci. USA} 115
(2018), 11730-11735.

$\left [16\right ]$
Cramer, J. G.. The Transactional Interpretation of Quantum Mechanics.\textit{ Rev. Mod.
Phys.} 58 (3) (1986), 647-688.

$\left [17\right ]$  Finkelstein, D. R.: Space-Time Sructure in High Energy Interactions. In Fundamental Interactions at High Energy, Gudehus, T., Kaiser, and G., A. Perlmutter G. A. editors. Gordon and Breach, New York 324-338 (1969). A pdf is available online at http://homepages.math.uic.edu/\symbol{126}kauffman/FinkQuant.pdf

$\left [18\right ]$  Feynman, R. P.: Simulating physics with computers. Int J. Theor. Phys. vol. 21  (1982), 467-488.

$\left [19\right ]$
Deutsch, D.: Quantum theory, the Church Turing principle and the
universal quantum computer. Proc. Roy.Soc. A 400,  (1985),
97-117.

$\left [20\right ]$ Ekert, A. K.: Quantum cryptography based on Bell's theorem. Phys. Rev. Lett. 67, 661 (1991).

$\left [21\right ]$
Rovelli, C.. Relational Quantum Mechanics. \textit{Int. J. Theor. Phys.} 35
(1996), 637-658.

$\left [22\right ]$
Fuchs, C. A.. On Participatory Realism. arXiv:1601.04360v3 [quant-ph] (2016).

$\left [23\right ]$
Long, G. L.. Grover algorithm with zero theoretical failure rate.
\textit{Phys. Rev. A} 64 (2001),            022307-022314.

$\left [24\right ]$
Toyama, F. M., van Dijk, W., and Nogami Y.. Quantum search with
certainty based on modified Grover algorithms: optimum choice of parameters.\textit{ Quant. Inf.
Proc.} 12 (2013), 1897-1914.

$\left [25\right ]$
Bennett, C. H., Bernstein, E., Brassard, G., and Vazirani, U..
Strengths and Weaknesses of Quantum Computing. \textit{SIAM Journal on Computing} 26
(1997), 1510-1523.

$\left [26\right ]$ Deutsch, D., Jozsa, R.. Rapid Solution of Problems by Quantum Computation. Proc. R. Soc. Lond. A 439 (1992), 553-55.

$\left [27\right ]$ Simon, D.. On the power of quantum computation. Proc. 35th Annual IEEE Symposium on the Foundations of Computer Science, (1994), 116-123.

$\left [28\right ]$ Shor, P. W.. Algorithms for quantum computation: discrete logarithms and factoring.  Proc. 35th Annual IEEE Symposium on the Foundations of Computer Science, (1994), 124-134.

$\left [29\right ]$ Mosca, M., Ekert, A.. The Hidden Subgroup Problem and Eigenvalue Estimation on a Quantum Computer.
QCQC 98, first NASA International Conference on Quantum Computing and Quantum Communication. Springer-Verlag London, UK (1998), 174-188.

$\left [30\right ]$
Price, H. and Wharton, K.. Disentangling the Quantum World. arxiv:1508.01140 (2015).

$\left [31\right ]$
Grover, L. K.. From Schrodinger equation to quantum search algorithm. arXiv:
quant-ph/0109116 (2001).

$\left [32\right ]$
Aaronson, S.. NP-complete problems and physical reality. \textit{ACM SIGACT
News}, 36(1) (2005), 30-52.

\end{document}